# The Impact of Industry Agglomeration on Land Use Efficiency: Insights from China's Yangtze River Delta

Hambur Wang

Guanghua School of Management, Peking University, Beijing 100871, China

## Abstract

This study investigates the impact of industrial agglomeration on land use intensification in the Yangtze River Delta (YRD) urban agglomeration. Utilizing spatial econometric models, we conduct an empirical analysis of the clustering phenomena in manufacturing and producer services. By employing the Location Quotient (LQ) and the Relative Diversification Index (RDI), we assess the degree of industrial specialization and diversification in the YRD. Additionally, Global Moran's I and Local Moran's I scatter plots are used to reveal the spatial distribution characteristics of land use intensification. Our findings indicate that industrial agglomeration has complex effects on land use intensification, showing positive, negative, and inverted U-shaped impacts. These synergistic effects exhibit significant regional variations across the YRD. The study provides both theoretical foundations and empirical support for the formulation of land management and industrial development policies. In conclusion, we propose policy recommendations aimed at optimizing industrial structures and enhancing land use efficiency to foster sustainable development in the YRD region.

## 1. Introduction

In the wave of 21st-century globalization, urban agglomerations have emerged as pivotal engines of regional economic development. The efficiency and intensity of land use within these urban clusters are critical to their potential for sustainable growth. The Yangtze River Delta (YRD), located on China's eastern coast, is endowed with significant geographical advantages and a robust industrial foundation, making it a key player in the nation's economic landscape. As global economic structures shift and China transitions its development model, the YRD faces the dual challenges of industrial upgrading and resource optimization. Since the early 2000s, the region has undergone rapid industrialization and urbanization. This accelerated growth in manufacturing and producer services has substantially increased the demand for land. Consequently, the urban land area in the YRD has expanded rapidly, highlighting concerns about the level of land use efficiency and intensification. Given its status as a major economic hub, the YRD's land use practices are not only crucial for regional economic health but also have far-reaching implications for national and global economic patterns. Ensuring efficient and intensive land use is essential for sustaining the region's economic vitality and addressing broader economic challenges.

Against this backdrop, industrial agglomeration has become a critical area of focus in both academic and policy research due to its significant influence on land use intensification. By optimizing resource allocation and enhancing production efficiency, industrial agglomeration contributes to more efficient and intensive land use within regions. This is particularly evident in the YRD, where the integration and concentration of manufacturing and producer services have not only driven rapid economic growth but also reshaped land use patterns. Studying the impact of industrial agglomeration on land use efficiency in the YRD is of both theoretical and practical significance. Theoretically, it expands the academic discourse on the interaction between industrial agglomeration and land use, providing new insights into coordinated industrial development and land resource management within urban agglomerations. Practically, the findings of this research can offer scientific support for land use policies and industrial development planning in the YRD. This can aid local governments in optimizing land management and industrial support strategies, thereby promoting sustainable economic development. Moreover, the policy recommendations derived from this study can serve as a reference for other regions aiming to achieve efficient land resource utilization and environmentally friendly economic growth.

This study focuses on the phenomenon of industrial agglomeration in the YRD (YRD) urban agglomeration and its impact on land use intensification. Utilizing data from 2003 to 2021, we systematically analyze the effects of manufacturing agglomeration, producer services agglomeration, and industrial synergy on urban land use intensification. Through theoretical analysis, we employ Ciccone and Hall's production density model to derive the mechanisms by which industrial agglomeration influences land use. Empirical tests are conducted using spatial econometric models and tools such as Stata, GeoDa, and Matlab to measure the degree of industrial agglomeration and the level of land use intensification. By deeply analyzing the characteristics of manufacturing and producer services agglomeration and their specific impacts on land use, this study aims to provide theoretical foundations and policy recommendations. These insights can aid government agencies and policymakers in optimizing land resource allocation and promoting rational industrial upgrading, thereby fostering sustainable and healthy economic development in the YRD and beyond. This not only enhances the efficiency of land resource utilization but also supports high-quality regional economic development.

## 2. Literature Review

### 2.1 Industrial agglomeration

The study of industrial agglomeration has a rich historical context, with foundational mechanisms and economic impacts extensively explored. The concept of spatial agglomeration in industries traces back to the late 19th century, when it was postulated that the provision of specific labor markets, technological spillovers, and the development of specialized intermediate goods and services drive agglomeration (Marshall, 1890). Moreover, the presence of scale economies and external economic effects were emphasized as direct drivers of industrial clustering. Further advancements in the early 20th century, such as Weber's (1909) analysis in "Theory of the Location of Industries," refined the geographic distribution of industrial agglomeration by highlighting the principle of cost minimization as a determinant in industrial location. By the late 20th and early 21st centuries, industrial agglomeration theories were expanded upon by scholars like Krugman (1991), who analyzed market demand, industrial localization, and external economies as primary reasons for industrial clustering based on theories of monopolistic competition and increasing returns to scale. Duranton and Puga (2004) examined the micro-foundations of urban agglomeration economies, proposing that mechanisms such as sharing of products and facilities, matching, and learning form the basis of urban clustering economies.

In more recent empirical studies, industrial agglomeration has been shown to significantly affect regional economies and business performance. For instance, concentrated industrial structures in the U.S. have been found to diminish economic performance, particularly impacting small businesses (Drucker and Feser, 2015). Moreover, the role of wholesale merchants in the early stages of industrial clustering was highlighted by Glasmeier (2015), who demonstrated how wholesaler functions facilitate greater local specialization and division of labor. Additionally, the relationship between industrial agglomeration and economic growth has been extensively explored. Studies have found that industrial clustering has a positive impact on regional economic growth, with significant regional heterogeneity observed in these effects (Xiang and Chen, 2017; Zhou, Li, and Li, 2016). In China, empirical analysis and policy application focusing on sectors such as manufacturing, service industries, and financial clustering have dominated the discourse. For example, analyses of manufacturing agglomeration and shifts in Guangdong Province highlighted the significant impact of internal economic factors on manufacturing mobility (Cao and Zhu, 2017). Financial services clustering was also studied, revealing significant local market roles in the repositioning of financial services across different regions (Zhang, 2016).

## 2.2 Intensive use of urban land

The study of intensive urban land use has its roots in agricultural land use research, with early inquiries focusing predominantly on agricultural contexts. The concept of "intensive" land use was originally defined by David Ricardo as the application of the same labor and capital inputs to a single piece of land. This foundational idea paved the way for more complex theories of urban land use. Urban land use models evolved significantly in the early 20th century with the development of three classic models: Burgess's (1923) concentric zone model, Hoyt and Homer's (1939) sector model, and Harris and Ullman's (1945) multiple nuclei model. These models integrated location theory with urban land use issues, offering a framework for understanding the dynamics of city growth and land use patterns (McMillen, 2003).

Recent international studies have shifted focus towards "compact" city models and "smart" urban growth, analyzing how these concepts affect the intensity and efficiency of land use. However, in China, the focus remains heavily on empirical exploration of intensive land use, including the definition of concepts, development of evaluative indices, and analysis of influencing factors. Among the notable Chinese studies, Tao (2000) argued that urban intensive land use should be based on rational layout, optimal land use structure, and sustainable development, enhancing land efficiency through better management and increased input of existing land resources. Xie, Hao, and Ding (2006) defined urban intensive land use as a dynamic and relative concept, emphasizing both economic and ecological benefits. Wang, Wang, and Jin

(2017) reiterated Ricardo's importance in defining the concept of land intensiveness, viewing it as a crucial starting point for discussions on maximizing output from existing land resources.

In terms of evaluation systems, recent significant contributions include Xie, Hao, and Ding's (2006) framework, which considers land input level, utilization intensity, and efficiency. Wang, Zhang, and Yao (2012) expanded this framework to include output benefits, utilization degree, and ecological quality. Sun, Lu, and Xiu (2015) proposed a system encompassing land input-output levels, regional linkage rationality, environmental capacity, and harmony between humans and land. Methodologically, techniques like principal component analysis, analytic hierarchy process (AHP), rank size rule, and fuzzy evaluation have been widely used to determine the weights of evaluation indices and to assess the intensity of urban land use. Zhu, Cui, and Miao (2017) utilized rank size rule methods to evaluate the intensity of land use across Chinese provinces, introducing time-serial three-dimensional data weighting that suits the dynamic nature of land use assessment. Liao (2018) employed an extendable stochastic environmental impact assessment model, STIRPAT, to empirically investigate the relationship between land use intensity and industrial structure, noting regional heterogeneity in linkage strength and the differential impact of demographic, policy, and protective factors on land use intensity across industries.

## 2.3 Relationship between industrial agglomeration and land use

The interplay between industrial agglomeration and urban land use has been extensively studied, with significant emphasis on how agglomeration impacts land use efficiency and intensification. Industrial agglomeration affects land use by modifying production efficiency as input densities increase, which in turn influences urban land intensification levels (Fujita and Thisse, 2002). Research indicates that local geographical externalities and the diversity of intermediate services can account for substantial variations in regional labor productivity, particularly by demonstrating how employment density correlates with productivity levels (Ciccone and Hall, 1996).

Further investigations have quantified the agglomeration effects in various European nations, revealing that while these effects exist, they are generally less pronounced than in the United States, with lower elasticity between labor productivity and employment density (Ciccone, 2000). In the Chinese context, studies utilizing panel data have explored how economic agglomeration and industrial structures influence urban land output rates, highlighting significant regional and city size heterogeneities in these impacts (Dou and Wang, 2010; Zhao and Jiang, 2013).

Another dimension of this research focuses on the productive service sector's role in urban land intensification. It has been argued that the clustering of productive services can lead to more intensive land use by promoting a shift in the structural use of industrial and service lands, thereby enhancing urban land use efficiency (Chu, 2013). Empirical studies employing spatial models have further validated the significant spatial correlations between productive service industry agglomeration and urban land intensification, noting that enhancing agglomeration levels can be beneficial for urban land use efficiency (Zhou and Tan, 2016). Moreover, spatial econometric analyses have shown that technological spillovers and scale economies progressively diminish their influence on land intensification levels, confirming the spatially heterogeneous effects of industrial agglomeration on urban land use (Cheng, 2017).

## 3. Methodology

### 3.1 Theoretical Analysis and Hypotheses

Industrial agglomeration fundamentally involves the concentration of essential production factors such as capital, labor, and technology within a specific geographic area. The resultant agglomeration effects influence regional industrial structures and the reallocation of resources, which are ultimately reflected in changes in land use patterns and intensification levels. Numerous studies have shown that the impacts of industrial agglomeration include both beneficial economies of agglomeration and potentially detrimental diseconomies of agglomeration.

**3.1.1 Agglomeration Economy and Intensive Land Use**

In contemporary studies of urban development and industrial economics, the interplay between industrial agglomeration and intensive land use has garnered significant attention. The economic impacts of agglomeration, particularly through spatial externalities, directly influence methods and efficiencies of urban land use.

Industrial agglomeration affects land intensification across three principal vectors: First, by fostering labor market pooling, agglomeration reduces recruitment and adjustment costs for enterprises, thereby enhancing production efficiency without additional land inputs (Marshall externalities). This pooling not only facilitates the congregation and optimization of specialized skills but also enhances collaborative and innovative capabilities within the industry, thus elevating the economic utility of land. Second, industrial agglomeration encourages the sharing of intermediate product markets. Proximity among different sectors reduces transportation and transaction costs and minimizes expenses related to storage and warehousing, thus promoting more effective land resource utilization (Marshall externalities). This shared market access for intermediate products not only lessens land occupation but also refines the structure of land inputs, boosting economic output per unit area. Third, industrial agglomeration catalyzes knowledge and technology spillovers. In agglomerated settings, firms can more readily share and propagate innovations, enhancing the adoption of new technologies and facilitating industrial upgrading, thereby directly altering the structure and efficiency of land use (Jacobs externalities). Particularly, the co-agglomeration of productive services and manufacturing speeds up the integration of complementary technologies and innovation, driving land use towards greater intensification and refinement.

Building on this theoretical framework, the following hypotheses are proposed to systematically evaluate the impact of industrial agglomeration on intensive land use:

**H1(a)**: Specialized agglomeration within the manufacturing sector will exert a positive impact on the level of intensive land use.

**H1(b)**: Specialized agglomeration within the productive services sector will exert a positive impact on the level of intensive land use.

**H1(c)**: Diversified agglomeration within the productive services sector will exert a positive impact on the level of intensive land use.

**H1(d)**: Synergistic agglomeration between the productive services and manufacturing sectors will exert a positive impact on the level of intensive land use.

### 3.1.2 Agglomeration Diseconomy and Intensive Land Use

This study explores the impact of industrial agglomeration on land use efficiency within the YRD urban agglomeration, drawing on theories of industrial agglomeration and land use intensification. Industrial agglomeration refers to the concentration of specific industries in a particular area, resulting in agglomeration effects. These effects can be divided into positive agglomeration economies and negative diseconomies of agglomeration. Agglomeration economies include economies of scale, external economies, and knowledge spillovers, where firms benefit from reduced production costs, shared resources, and knowledge, thus enhancing productivity and innovation. However, when the costs of agglomeration outweigh the benefits, diseconomies of agglomeration emerge, leading to a decline in land use efficiency.

Firstly, considering the changes in factor costs, the expansion of production scales due to industrial agglomeration leads to a continuous increase in the costs of labor, capital, and land. The scarcity and limited supply of land resources make land use costs particularly significant. During the conversion of agricultural land to industrial land and industrial land to service land, the high costs of land development and redevelopment rapidly increase the land use costs for firms in agglomeration areas, thereby reducing land use efficiency. Secondly, from the perspective of industrial location choice and bid-rent theory, the state of regional land use efficiency is closely related to industrial agglomeration. Factors determining the location of industrial agglomerations include transportation conditions, market demand, land rent, and firms' bidding ability. Due to the high output value-added and marginal revenue capability of productive services, they tend to agglomerate in central business districts (CBDs) or sub-centers. In contrast, manufacturing, with its large land scale and weak land rent payment ability, is typically located in more peripheral areas. The scarcity of land resources causes urban land prices to rise from the outskirts to the center, increasing land use costs in central agglomeration areas and suppressing land use efficiency. Lastly, considering the environmental costs of land use, industrial agglomeration can exert pressure on regional environments, resulting in issues such as insufficient residential land per capita, traffic congestion, reduced green coverage, and decreased ecological quality. The agglomeration of high-consumption and high-pollution manufacturing exacerbates regional pollution, further reducing land use efficiency.

Based on the theoretical analysis, this study proposes the following hypotheses:

**H2(a)**: The specialization agglomeration of manufacturing has a positive impact on land use efficiency.

**H2(b)**: The specialization agglomeration of productive services has a positive impact on land use efficiency.

**H2(c)**: The diversification agglomeration of productive services has a positive impact on land use efficiency.

**H2(d)**: The co-agglomeration of productive services and manufacturing has a positive impact on land use efficiency.

Additionally, considering that the impact of industrial agglomeration and co-agglomeration on land use efficiency may not be a simple linear relationship, the study further proposes the following nonlinear impact hypotheses:

**H3(a)**: The specialization agglomeration of manufacturing has a nonlinear impact on land use efficiency.

**H3(b)**: The specialization agglomeration of productive services has a nonlinear impact on land use efficiency.

**H3(c)**: The diversification agglomeration of productive services has a nonlinear impact on land use efficiency.

**H3(d)**: The co-agglomeration of productive services and manufacturing has a nonlinear impact on land use efficiency.

### 3.1.3 Theoretical Model Deduction

In examining the theoretical impacts of industrial agglomeration on land-use intensity, we utilize the output density models proposed by Ciccone and Hall (1993), as well as the extended model by Ushifusa and Tomohara (2013). Industrial agglomeration within a region increases the economic output density per unit of space, a concept theoretically validated by Marshall, Jacobs, and Porter, who demonstrated the external economic effects of agglomeration. Ciccone and Hall further explained the externalities of agglomeration from an output density perspective using a normative mathematical model. To elucidate the theoretical mechanisms through which industrial agglomeration influences urban land use, we rely on these models to conduct a detailed theoretical analysis.

The C-H model assumes a uniform spatial distribution of non-agricultural industries. The basic form of this model is as follows:

$$F(s_i, P_i, L_i) = P_i s_i^\alpha L_i^{1-\lambda}$$

In this equation, $F(s_i, P_i, L_i)$ represents the production density function; $s$ denotes the set of inputs required for production; $P_i$ is the total output of city $i$; $L_i$ indicates the land input area of city $i$. Thus, $P_i/L_i$ represents the land output density, or the output per unit area of land, which can also be understood as land-use efficiency or the level of land utilization. The parameter $\lambda$ indicates the output density coefficient, where $\lambda(1 - \lambda)$ represents the externality of agglomeration. When $\alpha > \lambda > 0$, agglomeration exhibits positive externalities, whereas when $\lambda < 0$, agglomeration shows negative externalities.

Ushifusa and Tomohara extended the basic C-H model by specifying labor and capital as the input factors, resulting in the following form:

$$p_i = \left(\frac{P_i}{L_i}\right) = \Omega_i \left(\frac{N_i}{L_i}\right)^{\beta_1} \left(\frac{K_i}{L_i}\right)^{\beta_2}$$

Here, $\Omega_i$ is the Hicks-neutral parameter, $N_i$ and $K_i$ represent labor and capital inputs, respectively, and $\beta_1$ and $\beta_2$ are their corresponding output elasticities. This model provides a more granular understanding of how specific input factors contribute to output density and, consequently, land-use efficiency.

## 3.2 Measurement and Analysis of Industrial Agglomeration

In measuring industrial agglomeration, we utilize the Location Quotient (LQ) to evaluate the degree of industrial specialization within a region. The LQ is calculated as follows:

$$LQ = \frac{\left(\frac{x_{ij}}{\sum_j x_{ij}}\right)}{\left(\frac{\sum_i x_{ij}}{\sum_i \sum_j x_{ij}}\right)}$$

where $x_{ij}$ represents the employment in industry $j$ in city $i$. The numerator indicates the share of employment in industry $j$ within city $i$, while the denominator represents the share of employment in industry $j$ nationwide. An LQ of 1 signifies an average level of specialization; an LQ greater than 1 indicates a higher concentration, while an LQ less than 1 suggests a lower concentration.

We focus on manufacturing and productive service industries, using employment data from key sectors. Additionally, the Relative Diversification Index (RDI) measures industrial diversification:

$$RDI_{ij} = 1 - \sum_j S_{ij}^2$$

where $S_{ij}$ is the employment share of industry $j$ in city $i$. A higher RDI indicates greater diversification, supporting Jacobs' externality theory.

Data from 2003 to 2021, covering 30 prefecture-level cities in the YRD Urban Agglomeration, is sourced from the "China Statistical Yearbook," "China City Statistical Yearbook," and provincial statistical bureaus. Missing data are interpolated as needed. This approach captures both specialized and diversified agglomeration of productive services while treating manufacturing data without further subdivision due to data constraints.

### 3.3 Vertical and horizontal pull grade evaluation method

In evaluating the intensive land use in urban areas, the method of differentiating vertical and horizontal tiers is essential. This approach addresses the need to simultaneously capture longitudinal and cross-sectional differences among the evaluated units. Based on the research by Zhu Z et.al (2017), this method is particularly suitable for the dynamic comprehensive evaluation of urban land use efficiency.

Now we will introduce a multi-period statistical analysis method to evaluate the production efficiency of multiple manufacturing units. Let there be $n$ manufacturing units, $S_1, S_2, \ldots, S_n$; each unit produces a product over $t_1, t_2, \ldots, t_n$ time periods. The variables $x_1, x_2, \ldots, x_m$ represent the different inputs for each period.

For each manufacturing unit and period, data is organized into matrices as shown in Table 1:

**Table 1: Data Organization for Multi-Period Efficiency Analysis**

| S | $t_1$<br>$x_1, x_2, \ldots, x_m$ | $t_2$<br>$x_1, x_2, \ldots, x_m$ | ... | $t_n$<br>$x_1, x_2, \ldots, x_m$ |
|---|---|---|---|---|
| $S_1$ | $x_1(t1), x_2(t1), \ldots, x_m(t1)$ | $x_1(t2), x_2(t2), \ldots, x_m(t2)$ | ... | $x_1(tn), x_2(tn), \ldots, x_m(tn)$ |
| $S_2$ | $x_1(t1), x_2(t1), \ldots, x_m(t1)$ | $x_1(t2), x_2(t2), \ldots, x_m(t2)$ | ... | $x_1(tn), x_2(tn), \ldots, x_m(tn)$ |
| ... | ... | ... | ... | ... |
| $S_n$ | $x_1(t1), x_2(t1), \ldots, x_{nm}(t1)$ | $x_1(t2), x_2(t2), \ldots, x_{nm}(t2)$ | ... | $x_1(tn), x_2(tn), \ldots, x_{nm}(tn)$ |

The efficiency analysis is performed using the following linear regression model for each unit and period:

$$y_i(t_k) = \sum_{j=1}^{m} w_j x_{ij}(t_k) \text{ for } i = 1,2,\ldots,n \text{ and } k = 1,2,\ldots,N$$

Where $y_i(t_k)$ represents the output in period $t_k$ for unit $i$, and $x_{ij}(t_k)$ represents the input $j$ for unit $i$ in period $t_k$. The weights $w$ represent the importance or impact of each input on the output.

The objective is to minimize the residual sum of squares, where the residual $v_i(t_k)$ is defined as:

$$v_i(t_k) = y_i(t_k) - \hat{y}_i(t_k)$$

The total sum of squared residuals is given by:

$$\sigma^2 = \sum_{k=1}^{N}\sum_{i=1}^{n} v_i(t_k)^2 = \sum_{k=1}^{N}(\mathbf{w}^T \mathbf{H}_k \mathbf{w}) - \sum_{k=1}^{N} \mathbf{w}^T \mathbf{H}_k \mathbf{w}$$

Where $\mathbf{H}_k = \mathbf{A}_k^T \mathbf{A}_k$ and $\mathbf{A}_k$ is the matrix of input values for all units in period $k$. The matrix $\mathbf{H}$ is the sum of all individual matrices $\mathbf{H}_k$ across periods.

Finally, the optimization aims to find the weight vector $\mathbf{w}$ such that the sum of squared residuals is minimized, subject to the constraint $\|\mathbf{w}\| = 1$, thus ensuring that the estimated weight vector maximizes the eigenvalue $\lambda_{max}(\mathbf{H})$. This process identifies the direction in input space that explains the greatest variance in output across all periods and units.

## 3.4 Establishment of the Spatial Panel Econometric Model

The spatial effects of urban land use have been well-documented in academic research, necessitating the application of spatial econometric models to avoid estimation biases. Spatial econometric methods, originally proposed by Anselin (1988), have evolved to include models such as the Spatial Lag Model (SLM) and the Spatial Error Model (SEM). This study employs both SLM and SEM, which are described in detail as follows:

### 3.4.1 Spatial Lag Model (SLM)

The SLM can be represented as:

$$y_{it} = \rho \sum_{j=1}^{n} \omega_{ij} y_{jt} + \sum_{k} \alpha_k x_{kit} + \sum_{m} \beta_m \text{ cont }_{mit} + \epsilon_{it}$$

In this model, $y_{it}$ is the dependent variable indicating the level of land use intensity in city $i$ at time $t$. The term $x_{kit}$ represents the key explanatory variables, which include the specialization agglomeration of manufacturing, the specialization agglomeration of producer services, the diversification agglomeration of producer services, and industrial synergy agglomeration. The term cont $_{mit}$ represents control variables, $\omega_{ij}$ is the standardized spatial weight matrix, using a geographical contiguity matrix where the elements are 1 if cities $i$ and $j$ are neighbors, and 0 otherwise. The parameter $\rho$ denotes the spatial autoregressive coefficient, indicating a positive spatial spillover effect when $\rho > 0$ and a siphon effect when $\rho < 0$.

### 3.4.2 Spatial Error Model (SEM)

The SEM is defined as:

$$y_{it} = \sum_{k} \alpha_k x_{kit} + \sum_{m} \beta_m \text{ cont }_{mit} + \mu_{it}$$

$$\mu_{it} = \lambda \sum_{j=1}^{n} \omega_{ij} \mu_{jt} + v_{it}$$

Here, the variables $y_{it}$, $x_{kit}$, cont $_{mit}$, and $\omega_{ij}$ retain the same meanings as in the SLM. The term $\mu_{it}$ represents the spatially correlated error component, with $\lambda$ as the coefficient of spatial autocorrelation of the error terms, and $v_{it}$ is the uncorrelated error term.

### 3.4.3 Selection of Control Variables

The selection of control variables is crucial to account for factors influencing urban land use intensity beyond the primary variables of interest. This study incorporates several control variables based on theoretical and empirical research:

(1) Economic Development Level (GDP): Measured by the gross domestic product per capita, reflecting the economic activity and wealth of the city.

(2) Land Resource Abundance (ABUND): Indicated by the proportion of undeveloped land to the total urban land area, representing the availability of land resources.

(3) Technological Development Level (TEC): Represented by the number of patent applications, indicating the technological innovation capacity of the city.

(4) Educational Level (EDU): Measured by the average years of schooling of the population, indicating human capital and labor quality.

(5) Urbanization Level (URBAN): Reflected by the urban population ratio, indicating the degree of urbanization.

(6) Industrial Structure Advancement (STR): Represented by the ratio of the tertiary industry to the secondary industry, indicating the advancement of the industrial structure.

By incorporating these control variables into the spatial econometric models, the analysis aims to isolate the specific impact of industrial agglomeration on urban land use intensity, ensuring a comprehensive and robust examination of the spatial dependencies and interactions within the Yangtze River Delta urban agglomeration.

## 4. Empirical Analysis

### 4.1 Analysis of industrial agglomeration in the YRD

#### 4.1.1 Producer service agglomeration

Producer services, initially distinguished from consumer services by Greenfield (1966), refer to industries providing goods or services to producers. These services, defined as knowledge-intensive sectors by Browning and Singelman (1975), include business, financial, and legal services. In China, Zhong and Yan (2005) characterized producer services as those catering to producers, government, or business activities without engaging in production or material transformation. According to the National Bureau of Statistics (2015), producer services encompass ten sectors, including R&D, information, leasing, and financial services. Despite varied research approaches, a consensus remains elusive, with scholars often selecting relevant sectors based on study needs. This study, aligning with Jiao and Lin's (2016) "five-industry" model, focuses on transportation, information technology, financial, leasing, and scientific research services.

Between 2003 and 2016, the average location quotient of producer services in the

YRD (YRD) was less than 1, indicating a lack of specialization and agglomeration on a national scale due to significant regional disparities. At a provincial level, Shanghai demonstrated a high degree of specialization with location quotients consistently above 1.5, reaching over 2 between 2013 and 2015, highlighting its role as a regional and national hub for producer services. Shanghai's advanced development in finance and information services, exemplified by districts such as Lujiazui and Zhangjiang Hi-Tech Park, underscores its economic prominence. Conversely, Jiangsu, Zhejiang, and Anhui provinces showed location quotients below 1, indicating underdeveloped producer services compared to Shanghai. Over time, both Jiangsu and Zhejiang displayed a declining trend in their location quotients, reflecting a lag in transitioning from traditional industrial structures to a more service-oriented economy. Anhui, with a simpler and smaller producer services sector, further illustrates the regional disparities.

In examining 30 cities across four selected years, a majority exhibited location quotients below 1, suggesting that most cities' producer services were underdeveloped relative to national averages, forming a "spindle-shaped" distribution. This pattern points to a nascent stage of agglomeration within the YRD, characterized by pronounced polarization. Shanghai stood out as the central hub, with provincial capitals Nanjing, Hangzhou, and Hefei also displaying location quotients above 1, forming a "one center and three sub-centers" model. The superior economic development and comprehensive industrial structures of these capitals facilitate their roles as regional service hubs. Notably, smaller cities like Zhoushan, Lianyungang, Lishui, and Quzhou maintained location quotients above 1 due to their inherent geographic and resource advantages, favoring the development of producer services over manufacturing. In contrast, economically advanced cities like Suzhou and Ningbo had lower location quotients for producer services, with Suzhou's economy driven by light industry and high-tech manufacturing, and Ningbo's by port-based industries.

From 2003 to 2016, the YRD's producer services sector showed that only the financial industry consistently maintained a location quotient above 1, indicating a comparative advantage at the regional level. The stable performance of the financial sector can be attributed to its foundational role in the economy and widespread branch network, which fosters workforce concentration. Other sectors, including transportation, information technology, leasing, and scientific research services, exhibited average location quotients below 1, with a declining trend, suggesting a move towards decentralization. Enhanced transportation networks across the YRD, reducing the relative advantage of traditional hubs like Shanghai and Nanjing, further illustrate this trend towards a more distributed agglomeration.

Analyzing the diversified agglomeration of producer services in the YRD (YRD) from both the urban agglomeration and provincial perspectives reveals significant insights. Due to differences in calculation methods, the values of specialization and diversification agglomeration are not directly comparable. Focusing on diversification, Jiangsu and Anhui provinces exhibit the highest levels, followed by Zhejiang, with Shanghai showing the lowest level of diversified agglomeration. The lower level in Shanghai is likely due to its high degree of specialization, particularly in the financial sector, which impedes diversification.

Over time, the diversified agglomeration levels in Shanghai, Jiangsu, and Zhejiang have shown a slow declining trend, while Anhui's trend remains unclear. The overall trend for the YRD urban agglomeration also lacks a clear pattern. At the city level, Figure 3.6 illustrates the average diversified agglomeration levels for 30 cities over the study period, arranged in ascending order. It reveals stark polarization, with Shanghai at the lowest end and Changzhou at the highest. The substantial manufacturing presence in Changzhou and Wuxi creates a demand for producer services, fostering diversified development. Although these cities may lack the total volume and specialization of producer services, the balanced development across various sectors and inter-sectoral interactions enhance their diversification. Other cities show smaller differences in diversified agglomeration levels, indicating relatively uniform development in producer services across the region.

### 4.1.2 Manufacturing agglomeration

**Table 2: Temporal Characteristics of Manufacturing Industry Agglomeration Levels in the Yangtze River Delta Urban Agglomeration from 2003 to 2021**

| Year | Shanghai | Jiangsu | Zhejiang | Anhui | Yangtze River delta region |
|------|----------|---------|----------|-------|----------------------------|
| 2003 | 1.367 | 1.484 | 0.988 | 1.170 | 1.221 |
| 2004 | 1.321 | 1.502 | 1.177 | 1.081 | 1.262 |
| 2005 | 1.166 | 1.517 | 1.301 | 1.039 | 1.291 |

| Year | | | | | |
|------|-------|-------|-------|-------|-------|
| 2006 | 1.237 | 1.543 | 1.326 | 1.027 | 1.307 |
| 2007 | 1.343 | 1.578 | 1.441 | 1.032 | 1.362 |
| 2008 | 1.346 | 1.603 | 1.504 | 0.994 | 1.380 |
| 2009 | 1.315 | 1.601 | 1.473 | 0.995 | 1.369 |
| 2010 | 1.291 | 1.633 | 1.451 | 1.016 | 1.377 |
| 2011 | 1.325 | 1.595 | 1.355 | 0.915 | 1.303 |
| 2012 | 1.407 | 1.603 | 1.259 | 0.926 | 1.280 |
| 2013 | 1.178 | 1.341 | 0.962 | 1.176 | 1.168 |
| 2014 | 1.029 | 1.364 | 1.033 | 1.153 | 1.183 |
| 2015 | 1.012 | 1.396 | 1.055 | 1.124 | 1.190 |
| 2016 | 1.054 | 1.416 | 1.093 | 1.148 | 1.218 |
| 2017 | 1.036 | 1.409 | 1.134 | 1.167 | 1.233 |
| 2018 | 1.043 | 1.255 | 1.282 | 1.255 | 1.282 |
| 2019 | 0.897 | 1.295 | 1.353 | 1.295 | 1.351 |
| 2020 | 0.905 | 1.306 |       | 1.306 | 1.370 |
| 2021 | 0.870 | 1.321 |       | 1.321 | 1.388 |
| Mean | 1.165 | 1.497 | 1.272 | 1.087 | 1.288 |

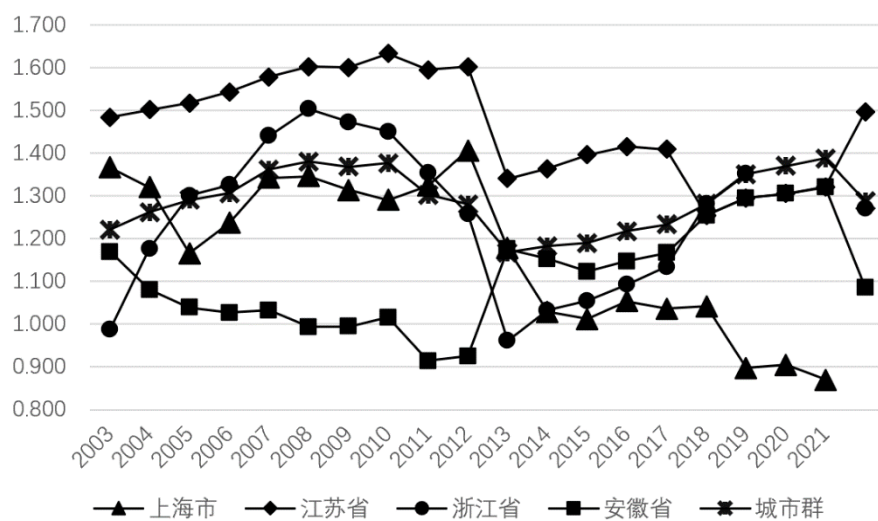

Between 2003 and 2016, the average location quotient of manufacturing in the YRD consistently exceeded 1, with an average level of 1.2, indicating a significant national comparative advantage and evident manufacturing agglomeration in the region. Analyzing Shanghai, Jiangsu, Zhejiang, and Anhui provinces reveals that Jiangsu has the highest manufacturing agglomeration, followed by Shanghai and Zhejiang, while Anhui lags. The slightly lower agglomeration level in Shanghai is due to its strategic shift towards service industries and the relocation of manufacturing to surrounding areas. Jiangsu and Zhejiang, with their historical and substantial manufacturing bases, have benefited from both the transfer of industries from Shanghai and successful foreign investment, enhancing their manufacturing clusters. Conversely, Anhui's lower manufacturing agglomeration results from its resource-based cities facing overcapacity and inefficiency issues, as well as a slower industrial transformation.

Examining specific cities within the YRD, most exhibit location quotients above 1, highlighting robust manufacturing agglomeration. However, significant disparities exist among cities. A new classification system is proposed to better capture these differences: cities with a location quotient above 2 form the first tier, those between 1.5 and 2 form the second tier, those between 1.0 and 1.5 form the third tier, and those below 1 form the fourth tier.

In the first-tier, cities like Suzhou and Jiaxing exhibit high manufacturing agglomeration. Suzhou, a pioneer of the "Southern Jiangsu Model," has maintained a manufacturing-focused economy post-reform, with a diverse range of industries and a substantial number of enterprises, including over 90 Fortune 500 companies by 2017. Jiaxing similarly focuses on manufacturing, leveraging innovation and new technologies for traditional industry upgrades. The second tier includes Wuxi, Changzhou, and Ningbo. Wuxi and Changzhou, benefiting from the "Southern Jiangsu Model," have successfully transitioned from traditional to precision and environmental manufacturing industries, maintaining high agglomeration. Ningbo, with its strong private sector foundation, has developed a comprehensive manufacturing ecosystem, boasting over 7000 large industrial enterprises and a robust industrial output. The third-tier features cities like Shanghai and Wuhu. Shanghai, despite its historical manufacturing prowess, has seen a decline in agglomeration due to strategic industrial restructuring towards services. Wuhu, an industrial hub in Anhui, has enhanced its manufacturing agglomeration by attracting industries relocating from coastal regions due to rising costs and environmental constraints. The 2010 "Wanjiang City Belt Industrial Transfer Demonstration Zone Plan" by the State Council further supports this trend. Finally, the fourth tier includes cities like Nanjing, Hefei, and Zhoushan. These provincial capitals serve more as political and cultural centers with diversified economies, emphasizing services over manufacturing. Zhoushan, constrained by its island geography, focuses on port-related industries and tourism, lacking the conditions for large-scale manufacturing development.

### 4.1.3 Producer service agglomeration

Analyzing the absolute differences in the agglomeration levels of producer services and manufacturing industries in the YRD from 2003 to 2016 reveals significant trends. Overall, the manufacturing sector in the YRD exhibits a higher agglomeration level compared to producer services, with location quotients consistently above 1 for manufacturing and below 1 for producer services, indicating a clear agglomeration in manufacturing while producer services are still in the nascent stages of clustering. The declining trend in the agglomeration of producer services can be attributed to the accelerated integration of the YRD, where capital and labor factors are flowing towards secondary central cities, leading to a more dispersed and diversified development rather than specialized clustering in producer services.

Table 3: Temporal Characteristics of Producer Services Agglomeration Levels in the Yangtze River Delta Urban Agglomeration from 2003 to 2021

| Year | Shanghai | Jiangsu | Zhejiang | Anhui | Yangtze River delta region |
|---|---|---|---|---|---|
| 2003 | 2.551399 | 1.397514 | 1.667748 | 1.215024 | 1.476257 |
| 2004 | 1.693375 | 0.92219 | 0.939429 | 0.868054 | 0.940459 |
| 2005 | 1.652197 | 0.909304 | 0.898562 | 0.868225 | 0.921066 |
| 2006 | 1.884424 | 0.885972 | 0.854856 | 0.840646 | 0.89915 |
| 2007 | 1.804746 | 0.845187 | 0.796942 | 0.804824 | 0.852685 |
| 2008 | 1.80236 | 0.83818 | 0.775636 | 0.821823 | 0.848196 |
| 2009 | 1.84209 | 0.803302 | 0.784767 | 0.858735 | 0.852022 |
| 2010 | 1.908737 | 0.788681 | 0.789401 | 0.845101 | 0.847122 |
| 2011 | 1.589404 | 0.811839 | 0.788319 | 0.816062 | 0.834049 |
| 2012 | 1.415384 | 0.837111 | 0.84327 | 0.816783 | 0.854559 |
| 2013 | 2.422488 | 0.718908 | 0.825222 | 0.847737 | 0.855613 |
| 2014 | 1.984903 | 0.696874 | 0.779 | 0.831688 | 0.811899 |
| 2015 | 1.977761 | 0.703865 | 0.757157 | 0.816419 | 0.80216 |
| 2016 | 1.929937 | 0.70417 | 0.763727 | 0.837784 | 0.809011 |
| 2017 | 1.905405 | 0.694566 | 0.816054 | 0.814863 | 0.815551 |
| 2018 | 1.908453 | 0.679269 | 0.787799 | 0.722567 | 0.7738 |
| 2019 | 1.823366 | 0.718532 | 0.801251 | 0.734655 | 0.791802 |
| 2020 | 1.823186 | 0.75405 | | 0.730094 | 0.805667 |
| 2021 | 1.892255 | 0.740075 | | 0.771726 | 0.822745 |
| Mean | 2.551399 | 1.397514 | 1.667748 | 1.215024 | 1.476257 |

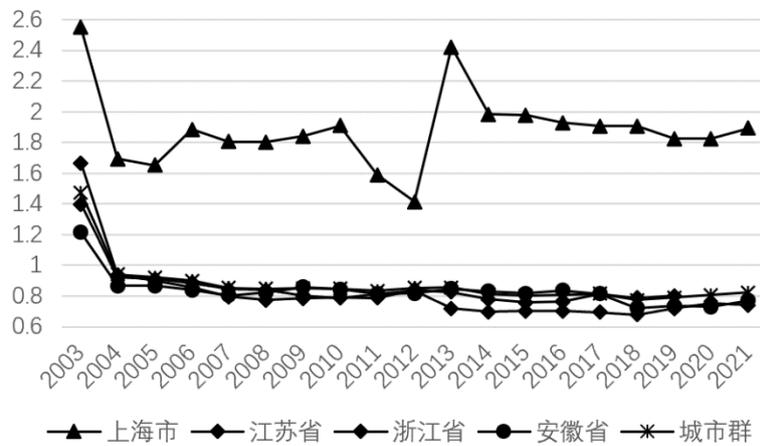

The absolute gap between the location quotients of the two industries shows an initial widening followed by a narrowing trend, indicating a fluctuating level of collaborative agglomeration between producer services and manufacturing. From an overall perspective, the collaborative agglomeration level between these industries in the YRD remains around 2.8 with little variation over the years.

From a provincial perspective, the collaborative agglomeration ranking is Shanghai > Jiangsu > Zhejiang > Anhui, with Shanghai significantly outperforming the other provinces. Within each province, significant disparities exist between cities, largely influenced by the respective agglomeration levels of producer services and manufacturing in each city. For instance, Shanghai tops the collaborative agglomeration index, reflecting its high-quality and extensive collaboration between the two sectors. In contrast, cities like Suqian in Jiangsu show much lower levels, highlighting the uneven development within provinces.

Table 4: Industry Synergy Agglomeration Index in the Yangtze River Delta Urban Agglomeration from 2003 to 2021

| Year | Shanghai | Jiangsu | Zhejiang | Anhui | Yangtze River delta region |
| --- | --- | --- | --- | --- | --- |
| 2003 | 2.923604 | 2.214979 | 1.896333 | 1.800175 | 2.012104 |
| 2004 | 2.387276 | 2.128072 | 1.769809 | 1.649209 | 1.876366 |
| 2005 | 2.15996 | 1.94535 | 1.795469 | 1.660536 | 1.825042 |
| 2006 | 2.326403 | 1.943906 | 1.738761 | 1.615033 | 1.792247 |
| 2007 | 2.289051 | 1.916012 | 1.823929 | 1.613991 | 1.809646 |
| 2008 | 2.310151 | 1.92204 | 1.816053 | 1.621461 | 1.812025 |
| 2009 | 1.904669 | 1.886868 | 1.779876 | 1.62737 | 1.774975 |
| 2010 | 2.65396 | 1.87313 | 1.767435 | 1.564226 | 1.775291 |
| 2011 | 2.304046 | 1.87669 | 1.736603 | 1.570122 | 1.754987 |
| 2012 | 1.866004 | 1.89174 | 1.736978 | 1.596666 | 1.75177 |
| 2013 | 2.26779 | 1.673685 | 1.695686 | 1.664358 | 1.700259 |
| 2014 | 2.151741 | 1.637584 | 1.676553 | 1.710507 | 1.691223 |
| 2015 | 1.886435 | 1.639457 | 1.640392 | 1.694587 | 1.665251 |
| 2016 | 2.132405 | 1.639029 | 1.642318 | 1.701232 | 1.676829 |
| 2017 | 2.378074 | 1.632382 | 1.651725 | 1.700296 | 1.68657 |
| 2018 | 2.34249 | 1.614741 | 1.661319 | 1.638613 | 1.664294 |
| 2019 | 1.912688 | 1.679516 | 1.72072 | 1.677908 | 1.70141 |
| 2020 | 2.118748 | 1.710288 | | 1.670677 | 1.715674 |
| 2021 | 2.240071 | 1.706712 | | 1.699135 | 1.73452 |
| Mean | 2.923604 | 2.214979 | 1.896333 | 1.800175 | 2.012104 |

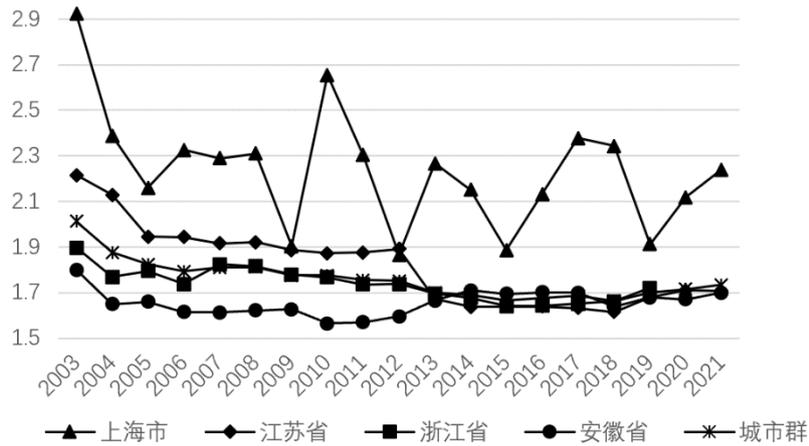

In Jiangsu, the disparity in collaborative agglomeration is stark, with Nanjing ranking second overall due to its high development level and strong industrial attraction, both manufacturing and producer services showing significant clustering. Conversely, Suqian ranks near the bottom, with most years showing location quotients below 1 for both sectors, indicating underdevelopment and low clustering. Similarly, Zhejiang displays a segmented distribution of collaborative agglomeration levels. Cities like Hangzhou, Jiaxing, Ningbo, and Quzhou rank within the top 10, benefiting from robust manufacturing bases that drive the development of producer services. Conversely, cities like Taizhou, Jinhua, Lishui, and Shaoxing, with weaker manufacturing foundations and slower transformation rates, show lower levels of collaborative agglomeration, remaining in the initial stages of development. In Anhui, collaborative agglomeration levels are uniformly low across cities, with minimal internal differences. Chuzhou consistently ranks last, dragging down the overall average for the province.

## 4.2 Evaluation of land intensive use of YRD

### 4.2.1 Construction of an Urban Land Intensive Use Evaluation Index System

The construction of an urban land intensive use evaluation index system primarily derives from the actual conditions of the evaluated area and existing standards, employing various selection methods. The first method is based on the PSR model, which evaluates pressure, state, and response. The second method selects indicators from perspectives such as land input intensity, land use intensity, economic benefits, social benefits, and sustainable development, which dominates academic literature.

The third method adheres to regulations and policy documents from land use departments, such as the "Evaluation Procedure for Intensive Use of Development Zone Land."

**Table 5: Land Intensive Utilization Level in the Yangtze River Delta Urban Agglomeration from 2003 to 2021**

| Year | Shanghai | Jiangsu | Zhejiang | Anhui | Yangtze River delta region |
| --- | --- | --- | --- | --- | --- |
| 2003 | 0.14464 | 0.102896 | 0.080542 | 0.08843 | 0.092705 |
| 2004 | 0.154368 | 0.105586 | 0.084551 | 0.090216 | 0.095827 |
| 2005 | 0.162482 | 0.115173 | 0.092311 | 0.092109 | 0.102471 |
| 2006 | 0.181783 | 0.128801 | 0.098632 | 0.088564 | 0.108785 |
| 2007 | 0.208827 | 0.145777 | 0.118238 | 0.0944 | 0.12371 |
| 2008 | 0.131955 | 0.152208 | 0.103214 | 0.092252 | 0.117362 |
| 2009 | 0.245131 | 0.160211 | 0.179656 | 0.114282 | 0.156229 |
| 2010 | 0.257036 | 0.147526 | 0.125617 | 0.116887 | 0.135201 |
| 2011 | 0.262377 | 0.123471 | 0.116061 | 0.090835 | 0.116476 |
| 2012 | 0.284997 | 0.158028 | 0.14311 | 0.132819 | 0.150289 |
| 2013 | 0.29959 | 0.16992 | 0.154694 | 0.139867 | 0.160743 |
| 2014 | 0.319557 | 0.177922 | 0.162609 | 0.144403 | 0.168132 |
| 2015 | 0.349476 | 0.352729 | 0.368693 | 0.27252 | 0.334164 |
| 2016 | 0.373701 | 0.188826 | 0.179794 | 0.151023 | 0.181462 |
| 2017 | 0.386239 | 0.183775 | 0.179098 | 0.157154 | 0.181827 |
| 2018 | 0.405819 | 0.20317 | 0.181848 | 0.167575 | 0.193021 |
| 2019 | 0.425693 | 0.20227 | 0.188306 | 0.176062 | 0.198125 |
| 2020 | 0.433186 | 0.206518 | 0.196974 | 0.181022 | 0.204178 |
| 2021 | 0.548384 | 0.231004 | 0.224937 | 0.202857 | 0.232397 |
| 均值 | 0.293434 | 0.171359 | 0.156783 | 0.136488 | |

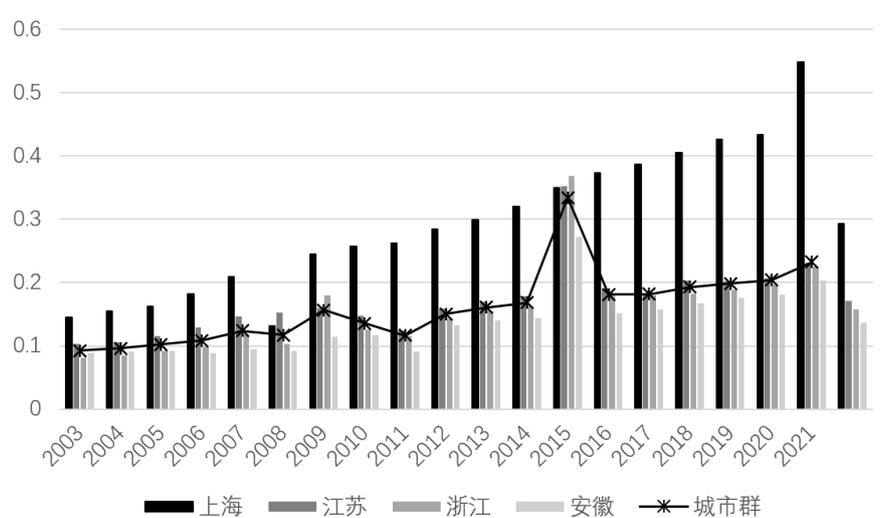

Considering the diversity and evolving nature of urban land use, the selection of evaluation indicators must be adaptable. This study, grounded in scientific, appropriate, systematic, and data-accessible principles, constructs an evaluation index system tailored to the YRD (YRD) urban agglomeration's unique context. Research indicates that the evaluation should focus on urbanization levels, economic development, and industrial structure of the region being assessed.

As one of the most urbanized regions in China, with an average urbanization level exceeding 65% in 2018, the YRD necessitates the inclusion of indicators such as population density (urban population/urban area) and per capita construction land (urban construction land area/urban population). Given the region's substantial economic output, with a GDP of 18 trillion yuan in 2018, economic impact indicators like per unit area fiscal revenue (total fiscal revenue/built-up area), per unit area GDP (total GDP/built-up area), and per unit area retail sales of consumer goods (total retail sales/built-up area) are crucial.

Thus, integrating these considerations and existing scholarly findings, the evaluation system for land intensive use in the YRD is constructed across three dimensions: land use intensity, economic benefits, and ecological benefits. The evaluation unit encompasses 30 prefecture-level cities from 2003 to 2016, with data sourced from the "China Statistical Yearbook," "China City Statistical Yearbook," the National Bureau of Statistics, and provincial and municipal statistical bureaus.

**4.2.2 Measurement and Analysis of Land Intensive Use Levels in YRD**

The evaluation of land intensive use levels in the YRD from 2003 to 2016 employs a method combining normalization of input indicators and calculation of their weights, resulting in a comprehensive measure of land use intensity across 30 prefecture-level cities. Time series analysis reveals that Shanghai consistently exhibits the highest level of land intensive use, far exceeding the regional average. Jiangsu follows closely, aligning with the regional average, while Zhejiang's land use intensity ranks third, catching up to Jiangsu post-2012. Anhui consistently shows the lowest levels, significantly below the regional average and less than half of Shanghai's levels

Shanghai's land use intensity demonstrates a "W-shaped" trend with significant fluctuations, peaking in 2003 and 2009 before declining. In contrast, Jiangsu, Zhejiang, Anhui, and the overall region display a more uniform trend of initial decline followed by an increase and subsequent decline, though with less variability than

Shanghai.

The significance lies in the relative ranking and temporal changes in land use intensity among cities. A dynamic analysis of maximum sequence differences categorizes cities based on stability and ranking. Cities are classified as stable if their maximum sequence difference is between 0 and 5, fluctuating between 6 and 10, and jumping if above 10. Additionally, cities are ranked based on land use intensity: highly intensive (rank 0-10), moderately intensive (rank 11-20), and low intensive (rank 21-30).

The analysis shows three cities with stable land use: Huainan, Huaian, and Nanjing. Both Huainan and Huaian consistently rank low in land use intensity, around 0.2, indicating "stable low intensity." Nanjing, ranking around 15th, falls under "stable moderate intensity." Thirteen cities exhibit fluctuating land use, mostly in Jiangsu. Cities like Suzhou, Taizhou, Nantong, and Wuxi, with rankings generally within the top 10, are "fluctuating high intensity," while Zhenjiang and Yangzhou, ranking mid-range, are "fluctuating moderate intensity." In Anhui, Maanshan, Hefei, and Wuhu, with rankings mostly post-20, are "fluctuating low intensity." Shanghai, despite high rankings, shows a fluctuating pattern due to its limited land resources and strategic policies promoting intensive land use and industrial restructuring. Fourteen cities, primarily in Zhejiang, exhibit jumping land use patterns. Cities like Chuzhou, Xuzhou, Wenzhou, Suqian, and Lianyungang show the most significant jumps and lowest rankings, classifying them as "jumping low intensity." Other cities display large, irregular changes in rankings, spanning across classification standards, indicating highly variable land use intensity.

Overall, Zhejiang's cities show the greatest variability in land use intensity, while Anhui's cities, predominantly fluctuating and low intensive, reflect the province's lagging development. Jiangsu's cities display a mix of stability, fluctuation, and jumping, generally maintaining moderate to high land use intensity, except for Nanjing, which is stably moderate. This analysis highlights the diverse and evolving patterns of land use intensity within the YRD.

## 4.3 Impact of industrial agglomeration on land intensive use

Considering the situation of multiple models, we finally choose the spatial fixed effect model under the spatial lag model SLM to conduct spatial econometric estimation of land intensive use in the YRD urban agglomeration.

## 4.3.1 Analysis of Manufacturing Agglomeration's Impact on Land Intensive Use

The empirical results indicate that the specialization agglomeration of manufacturing in the YRD (YRD) positively affects land intensive use, validating hypothesis H1(a) while rejecting H2(a). The non-significant results for the quadratic term suggest no nonlinear impact, thus rejecting H3(a). During the study period, the average level of manufacturing agglomeration in the YRD was around 1.2, with most cities showing values above 1, reflecting a strong clustering trend in manufacturing. Manufacturing remains a key economic driver, leveraging land more intensively for economic growth through positive externalities and ongoing agglomeration economies.

Internally, the concentration of manufacturing firms enhances product, factory, and company specialization, increasing returns to scale and thus land use efficiency. Externally, the clustering of vertically or cost-linked firms reduces storage needs for raw materials and finished goods, saving land and boosting land use efficiency. Urbanization effects further amplify these benefits by attracting complementary industries and services, increasing population and infrastructure density, and thus enhancing land input intensity and usage efficiency.

Regarding control variables, the positive coefficient for GDP indicates that higher economic development levels enhance land intensive use, aligning with findings that cities like Shanghai have higher land use efficiency compared to less developed areas like Anhui. Land resource abundance shows a negative impact, confirming the "land resource curse," where cities with more land resources tend to use them less efficiently due to poor planning and underutilization. Technological development positively influences land intensive use, with increased patent activity reflecting higher innovation and productivity, driving efficient land use. Higher education levels also positively affect land intensive use by improving labor quality and productivity. Urbanization enhances land intensive use, supporting the concentration of population and industries, which boosts land efficiency. However, the degree of industrial structure sophistication (STR) does not significantly affect land intensive use, suggesting that the proportion of tertiary to secondary industries does not drive land use efficiency during the study period.

Table 6: Spatial Econometric Results of the Impact of Manufacturing Specialization Agglomeration on Land Intensive Utilization

| VARIABLES | (1) y | (2) y |
| --- | --- | --- |
| LQagman | 0.020*** | 0.041 |
|  | (3.05) | (1.48) |
| LQagman2 |  | -0.007 |
|  |  | (-0.78) |
| GDP | 0.000*** | 0.000*** |
|  | (8.76) | (8.78) |
| ABUND | -0.100** | -0.104** |
|  | (-2.19) | (-2.25) |
| TEC | -0.001*** | -0.001*** |
|  | (-5.96) | (-5.62) |
| EDUC | 0.002 | 0.002 |
|  | (0.44) | (0.60) |
| URBAN | 0.106*** | 0.099*** |
|  | (3.75) | (3.31) |
| STR | 0.078*** | 0.078*** |
|  | (7.45) | (7.47) |
| Constant | -0.056 | -0.072* |
|  | (-1.62) | (-1.79) |
| Observations | 474 | 474 |
| R-squared | 0.566 | 0.566 |

t-statistics in parentheses
*** p<0.01, ** p<0.05, * p<0.1

## 4.3.2 Analysis of Producer Services Agglomeration's Impact on Land Intensive Use

The specialization agglomeration of producer services in the YRD (YRD) has a negative impact on land intensive use, supporting hypothesis H2(b) and rejecting H1(b). The lack of significance in the quadratic term coefficients indicates no nonlinear impact, thus rejecting H3(b). During the study period, the overall level of specialization agglomeration in producer services in the YRD was 0.89, with values below 1 in Jiangsu, Zhejiang, and Anhui, except for Shanghai. The agglomeration levels of the five sub-sectors, excluding finance, were also below 1, indicating a lack of specialization agglomeration at the national level. Despite the significant role of producer services in economic development, their lack of scale economies leads to a negative impact on land intensive use. According to location and rent theories, producer services are concentrated in urban centers where high land development and redevelopment costs elevate land use costs. Additionally, significant manufacturing agglomeration in the YRD squeezes producer services spatially and resource-wise, slowing their agglomeration. Cities in the YRD tend to develop all five sub-sectors of producer services uniformly, preventing any single sector from achieving significant agglomeration advantages. Thus, high land costs, the manufacturing squeeze, and diversified urban development hinder producer services from forming specialized agglomerations, leading to low agglomeration economies, and negatively affecting land intensive use.

Producer services diversification agglomeration, however, shows a nonlinear "inverted U-shaped" effect on land intensive use, supporting hypothesis H3(c) while rejecting H1(c) and H2(c). This aligns with the view that diversified rather than specialized agglomeration promotes economic growth. The high diversification level indicates significant Jacobs externalities, emphasizing the economic effects of different industries clustering in a region due to knowledge and technology spillovers and urban scale economies. Developed manufacturing in the YRD fosters producer services growth, yet the limited space and slow industrial transition hinder large-scale specialization agglomeration, leading to high diversification. Diversified producer services enhance knowledge and technology spillovers, improving land development technology and intensive land use. Technological collisions within diversified agglomerations lead to new innovations like fintech, advancing industrial structure and land use optimization. However, as with specialization, the economic effects of diversification are not always positive. When the costs of agglomeration exceed the benefits, issues like high land rents, congested spaces, and diminishing returns arise, reducing external economies and leading to dispersed layouts. This "inverted U-shaped" impact of diversification agglomeration on land intensive use during the study period indicates both economic and diseconomy behaviors of agglomeration.

**Table 7: Spatial Econometric Results of the Impact of Producer Services Specialization Agglomeration on Land Intensive Utilization**

| VARIABLES | (1) y | (2) y |
|---|---|---|
| LQagser | -0.003 | 0.002 |
|  | (-0.26) | (0.08) |
| LQagser2 |  | -0.002 |
|  |  | (-0.18) |
| GDP | 0.000*** | 0.000*** |
|  | (8.22) | (8.22) |
| ABUND | -0.071 | -0.071 |
|  | (-1.35) | (-1.35) |
| TEC | -0.001*** | -0.001*** |
|  | (-5.21) | (-5.21) |
| EDUC | 0.000 | 0.000 |
|  | (0.11) | (0.11) |
| URBAN | 0.119*** | 0.120*** |
|  | (4.21) | (4.21) |
| STR | 0.063*** | 0.064*** |
|  | (6.20) | (6.18) |
| Constant | -0.018 | -0.021 |
|  | (-0.55) | (-0.58) |
| Observations | 474 | 474 |
| R-squared | 0.557 | 0.557 |

t-statistics in parentheses
*** $p<0.01$, ** $p<0.05$, * $p<0.1$

### 4.3.3 Analysis of the Impact of Industrial Co-agglomeration on Land Intensive Use

The empirical analysis reveals that the collaborative agglomeration of producer services and manufacturing in the YRD (YRD) has a nonlinear "inverted U-shaped" effect on land intensive use. This confirms hypothesis H3(d) while rejecting H1(d) and H2(d). The results indicate that the initial positive effects of collaborative agglomeration on land use efficiency eventually turn negative as the level of agglomeration increases, aligning with the dual nature of agglomeration effects proposed by researchers like Li Qiang and Dou Jianmin.

In the YRD, the strong manufacturing base initially drives the demand for complementary producer services, leading to significant external economies such as shared labor markets, intermediate goods markets, and knowledge spillovers. This mutual reinforcement between manufacturing and producer services promotes specialized and diversified agglomerations. For instance, the integration of internet technologies into manufacturing has spurred the development of e-commerce and smart manufacturing, enhancing overall productivity, and enabling economic growth through increased land use efficiency.

During the initial stages of collaborative agglomeration, the positive spillover effects dominate, as expanding land use and increasing land inputs contribute to higher productivity and economic returns. The transition from extensive to intensive land use, supported by relatively low land prices and increasing returns to scale, facilitates this growth. Enhanced cooperation and technological spillovers between industries further drive this phase.

However, as collaborative agglomeration intensifies, the positive effects give way to congestion effects. The finite nature of land resources becomes increasingly restrictive, leading to higher land costs and diminishing returns. The relationship between producer services and manufacturing shifts from cooperative to competitive, with the industries potentially crowding each other out. The scarcity of developable land, rising input costs, and reduced economic outputs from land use contribute to this phase. High agglomeration costs may force some firms to relocate, causing industrial dispersion and diminishing agglomeration economies. Consequently, land intensive use declines, reverting from intensive to more extensive land use patterns.

Table 8: Spatial Econometric Results of the Impact of Industry Synergy Agglomeration on Land Intensive Utilization in the Yangtze River Delta Urban Agglomeration

| VARIABLES | (1) Score | (2) Score |
|---|---|---|
| Cogg | 0.023** | 0.131** |
|  | (2.29) | (2.25) |
| Cogg2 |  | -0.030* |
|  |  | (-1.88) |
| GDP | 0.000*** | 0.000*** |
|  | (9.17) | (9.05) |
| ABUND | -0.127** | -0.132*** |
|  | (-2.53) | (-2.62) |
| TEC | -0.001*** | -0.001*** |
|  | (-5.37) | (-5.47) |
| EDUC | 0.001 | 0.002 |
|  | (0.15) | (0.54) |
| URBAN | 0.106*** | 0.105*** |
|  | (3.68) | (3.65) |
| STR | 0.064*** | 0.065*** |
|  | (6.98) | (7.04) |
| Constant | -0.055 | -0.161** |
|  | (-1.52) | (-2.40) |
| Observations | 474 | 474 |
| R-squared | 0.562 | 0.565 |

t-statistics in parentheses
*** p<0.01, ** p<0.05, * p<0.14.3.1

## 5. Results and Discussion

### 5.1 Main Findings

The study's conclusions encompass several key findings based on extensive literature review and empirical analysis of industrial agglomeration, land intensive use, and their interrelations. The research involved theoretical analysis and empirical investigation using the production density model, focusing on the impact of industrial agglomeration and industrial collaborative agglomeration on land intensive use.

First, the study calculated the specialization agglomeration of manufacturing, specialization agglomeration of producer services, diversification agglomeration of producer services, and industrial collaborative agglomeration in the YRD (YRD) from 2003 to 2016. The results indicated that the YRD exhibited significant manufacturing agglomeration, especially in Jiangsu, Zhejiang, and Shanghai. However, the specialization agglomeration of producer services was generally below 1, except in Shanghai, indicating a lack of significant clustering. Diversification agglomeration of producer services was highest in Jiangsu and Anhui, followed by Zhejiang and Shanghai. The industrial collaborative agglomeration remained relatively stable over the years, with Shanghai showing the highest levels.

Second, the evaluation of land intensive use in the YRD revealed that Shanghai had the highest levels, followed by Jiangsu and Zhejiang, with Anhui trailing below the regional average. The temporal trend showed the most fluctuation in Shanghai, while the other regions exhibited a more consistent pattern. Cities like Huainan, Huaian, and Nanjing displayed stable land use levels, whereas many cities in Jiangsu showed fluctuating patterns. The spatial distribution of land intensive use highlighted clear clustering, with higher levels in coastal and southern areas and lower levels in the northwest.

Third, spatial lag panel models were employed to estimate the impact of industrial agglomeration on land intensive use. The results demonstrated positive spatial autocorrelation, indicating significant clustering of land intensive use across neighboring cities. Manufacturing specialization agglomeration had a positive impact without nonlinear effects, suggesting ongoing significant economic benefits from agglomeration. In contrast, the specialization agglomeration of producer services negatively impacted land intensive use, likely due to insufficient clustering. The diversification agglomeration of producer services and industrial collaborative agglomeration both exhibited "inverted U-shaped" effects, highlighting the dual nature of agglomeration economies and diseconomies.

Control variables such as economic development, technological advancement, education levels, and urbanization positively influenced land intensive use, while land resource abundance had a negative impact. The degree of industrial structure sophistication did not show a significant effect. The robustness of the model was confirmed by consistent results when land intensive use levels estimated by the entropy method were reanalyzed.

Overall, the study underscores the complex dynamics of industrial agglomeration and its varying impacts on land intensive use in the YRD, highlighting the need for balanced development strategies to maximize the benefits of agglomeration while mitigating its negative effects.

## 5.2 Policy Implications

The YRD (YRD) should emphasize the continuous development of manufacturing, particularly high-end manufacturing, to drive and promote the further growth of producer services. Maintaining the development advantage of manufacturing within the region is crucial. Key cities like Suzhou, Wuxi, and Ningbo are major manufacturing hubs, and the significant role of manufacturing in driving regional economic stability and growth should not be overlooked. The scale of manufacturing agglomeration remains evident and positively impacts land use and economic development. Efforts should focus on leveraging the integration of the YRD to facilitate the transfer of manufacturing from core cities like Shanghai, Suzhou, and Ningbo to less developed areas such as Anhui, thus accelerating the development of non-core cities.

Enhancing high-end manufacturing within the region is also essential. Innovation is fundamental to a city's progress, and cities in the YRD should align with the "Made in China 2025" initiative, focusing on developing high-end manufacturing. Shanghai should continue to advance its industrial transformation, creating a world-class high-end manufacturing center, while cities like Hangzhou, Nanjing, and Hefei should welcome the spillover of technology, talent, and capital from Shanghai. This would form a development pattern where provincial capitals drive high-end manufacturing in other cities within their provinces. Additionally, cities with robust manufacturing should systematically phase out low-end manufacturing to climb higher in the global manufacturing value chain.

Promoting the agglomeration of related producer services alongside manufacturing is vital. This involves fostering both specialization and diversification in producer services to enhance scale economies and technological spillover effects. Although the specialization agglomeration of producer services currently does not significantly enhance land use efficiency, this should not deter efforts to develop these services. The goal should be to establish a networked development pattern centered on Shanghai, with Nanjing, Hangzhou, and Hefei as secondary centers, promoting the growth and sophistication of producer services in alignment with each city's development needs (Wang & Su, 2024). Research shows that diversified agglomeration enhances knowledge and technology spillovers, which are critical for regional economic growth (Huo et al., 2024).

Cities should scientifically plan the collaborative layout of manufacturing and producer services, cautiously advancing the "dual-engine" development strategy. The study highlights that manufacturing location effects, through industrial linkage mechanisms, attract producer services, emphasizing the need for a scientifically planned and rationally laid out collaborative development of these industries within the region. The dual nature of collaborative agglomeration effects suggests that while it fosters technological exchanges and economic efficiency initially, excessive agglomeration can lead to congestion and resource misallocation. Therefore, before implementing the dual-engine strategy, cities should conduct detailed investigations into the specialization agglomeration of both industries to devise region-specific strategies that enhance the depth and quality of industrial collaboration (Yao et al., 2024).

Lastly, land use policies should be tailored to local conditions to enhance land intensive use. The positive spillover effects of land intensive use between neighboring cities in the YRD should be leveraged to promote coordinated and integrated development. High-intensity land use cities should radiate their practices to surrounding areas, fostering a region-wide pattern of high land use efficiency (Yao et al., 2024). Research on time-sequence tracking technologies, like FMRFT, can provide innovative solutions for land and resource monitoring (Yao et al., 2024). Moreover, new technologies, such as self-supervised learning models applied in agricultural monitoring, play a crucial role in enhancing land and environmental management (Wang et al., 2024).

## 6. Conclusion

This study provides a comprehensive analysis of the impact of industrial agglomeration on land use efficiency in the YRD (YRD), utilizing spatial econometric models and a range of quantitative measures. The findings underscore the complex and multifaceted nature of industrial agglomeration effects on land use, highlighting both the benefits and challenges associated with these phenomena.

The empirical results confirm that manufacturing specialization agglomeration positively influences land intensive use, driven by significant economies of scale and enhanced productivity from concentrated industrial activities. The clustering of manufacturing firms leads to more efficient land use through improved production processes, reduced storage needs, and optimized land resource allocation. These positive externalities underscore the importance of maintaining and enhancing the manufacturing base within the YRD to sustain economic growth and land use efficiency.

Conversely, the specialization agglomeration of producer services exhibits a negative impact on land intensive use, reflecting the sector's current lack of sufficient clustering and scale economies. High land development costs in urban centers, coupled with the spatial competition from dominant manufacturing sectors, hinder the effective agglomeration of producer services. This finding suggests a need for targeted policy interventions to foster the growth and clustering of producer services, potentially through incentives and strategic planning that align with urban development goals.

The study also reveals an inverted U-shaped relationship between the diversification agglomeration of producer services and land use efficiency. While initial diversification promotes technological spillovers and innovation, leading to more efficient land use, excessive agglomeration results in congestion effects, rising land costs, and diminished returns. This dual nature of diversification agglomeration calls for a balanced approach that encourages sectoral diversity without overburdening urban land resources.

Furthermore, the collaborative agglomeration of manufacturing and producer services shows a similar inverted U-shaped effect on land use efficiency. Initial synergies between these sectors enhance productivity and land use through shared resources and knowledge spillovers. However, as agglomeration intensifies, competition for limited land resources and increasing costs reduce these benefits, highlighting the need for careful management of industrial collaborations.

Overall, the study emphasizes the importance of strategic planning and policy formulation to maximize the benefits of industrial agglomeration while mitigating its negative impacts. This includes promoting high-end manufacturing, fostering the growth of producer services, and ensuring a balanced and coordinated development approach that leverages the strengths of both sectors. Tailored land use policies that consider regional characteristics and development stages are crucial for sustaining the economic vitality and environmental sustainability of the YRD. The insights and recommendations provided by this research offer valuable guidance for policymakers and urban planners aiming to optimize land resource utilization and support sustainable economic development in the YRD and similar urban agglomerations globally.